# A simple and rapid method for detecting living microorganisms in food using laser speckle decorrelation

Jonghee Yoon[1, 3, +], KyeoReh Lee[1, 4, +], and YongKeun Park[1,2,*]

[1] Department of Physics, Korea Advanced Institutes of Science and Technology, Daejeon 34141, South Korea

[2] TOMOCUBE, Inc., Daejeon 34141, South Korea

[3] jh.yoon@kaist.ac.kr

[4] kyeo@kaist.ac.kr

+These authors are contributed equally to this work

*Corresponding author, E-mail: yk.park@kaist.ac.kr, Phone: +82-42-350-2514, Fax: +82-42-350-2510

**Abstract**

Measuring microorganisms in food products is a critical issue for food safety and human health. Although various approaches for detecting low-levels of microorganisms in food have been developed, they require high-cost, complex equipment, invasive procedures, and skilled technicians which limit their widespread use in the food industry. Here, we present a simple, non-destructive, non-contact, and rapid optical method for measuring living microorganisms in meat products using laser speckle decorrelation. By simply measuring dynamic speckle intensity patterns reflected from samples and analyzing the temporal correlation time, the presence of living microorganisms can be non-invasively detected with high sensitivity. We present proof-of-principle demonstrations for detecting *E. coli* and *B. cereus* in chicken breast tissues.

**Keywords**

**Introduction**

The spoilage and poisoning of food products by microorganisms are crucial issues in food safety and human health. The growth and activity of various types of microorganisms, including bacteria, yeast, and mold in food deteriorate food quality and cause food intoxication.[1,2] To detect and investigate microorganisms and pathogenic agents in food products, many methods, including microbiological culture methods,[3] high-performance liquid chromatography,[4] hyperspectral imaging,[5] Raman spectroscopy,[6] nuclear magnetic resonance technique,[7] and mass spectroscopy,[8] have been developed. These methods have contributed to ensuring food safety and quality by providing information about the presence of pathogenic microorganisms. However, these conventional methods require laboratories, high-cost equipment, and professionals. Moreover, complicated procedures for sample preparation and long analysis times are needed thus preventing rapid detection and implementation in white electronics found in stores and homes. The limitations above have restricted their widespread use in food processing, transportation, marketing, and preservation in various food industries.

In this study, we propose and experimentally demonstrate a simple, non-destructive and rapid optical method for measuring the quality of meat products using laser speckle imaging. Laser speckle imaging has been introduced to monitor moving particles in optically inhomogeneous media by analyzing time-varying laser speckle patterns.[9,10] Light impinging on turbid media such as biological tissues experiences multiple light scattering, and scattered light produces laser speckle patterns by light interference.[11] Because of the deterministic nature of multiple light scattering, scattered light from a static turbid medium generates a constant laser speckle pattern.[12] However, if scatters are spontaneously moving inside a turbid medium, time-varying speckle patterns are produced, from which information of the moving scatters can be retrieved.[9,10,13] Previously, laser speckle imaging has been applied to visualize blood flow in the retina[14] or brain[10,15] and food conditions such as seed viability,[16] shelf life of fruits,[17,18] aging of beef[19] and muscle properties in meat.[20]

In this work, we propose and demonstrate that measuring the dynamic speckle patterns from meat enables the detection of any living microorganisms present. Unlike multiple light scattering in meat which exhibits static and deterministic speckle intensity patterns, light paths associated with the movements of living microorganisms result in time-varying changes in the speckle intensity patterns. Thus, by detecting the decorrelation in the laser speckle intensity patterns from tissues, the living activities of microorganisms can be detected. As proof-of-principle demonstrations, we present the detection *E. coli* and *B. cereus* in chicken breast tissues by measuring the maps of the speckle decorrelation time.

**Materials and Methods**

**Optical setup**

The experimental setup is shown in Figure 1b. A coherent laser ($\lambda$ = 633 nm, 5 mW, HRR050-1, Thorlabs Inc., USA) is used as an illumination source. The laser beam is illuminated onto a sample, and scattered intensity images are captured with a CCD camera (Lt365R, Lumenera Inc., USA). The speckle grain size, which is determined by the numerical aperture of an imaging system, was 23 μm at the camera plane, which corresponds to 5 × 5 camera pixels. Dynamic speckle images were obtained at a frame rate of 30 Hz for 20 seconds.

Even though we used an optical imaging system to systematically capture the speckle intensity patterns, the implementation of the present method is not limited to the system shown in Figure 1b and can be achieved with a general optical system consisting of coherent illumination and a detection device. For example, the present method can also be implemented in an existing refrigerator with a laser diode and a simple photodetector.

**Bacterial sample preparation**

Two bacterial species, *Escherichia coli* (*E. coli*, 22002, Korean Collection for Type Cultures, Republic of Korea) and *Bacillus cereus* (*B. cereus*, 1092, Korean Collection for Type Cultures, Republic of Korea) were used in the experiment. These are representative bacterial species that can cause food poisoning[21]. *E. coli* was cultured with Luria-Bertani (LB) broth media and LB agar plates (244620, BD Bioscience, USA), and *B. cereus* was cultured with Nutrient broth media and Nutrient agar plates (234000, BD Bioscience, USA). Diluted cells were grown in 5 mL of growth media in 15 mL conical tubes for 12 h in a 37°C incubator with vigorous shaking (150 rpm). Bacterial concentrations in the suspensions were calculated by measuring the light absorbance at 600 nm with a spectrophotometer (SpectraMax Plus 384, Molecular Devices Inc., USA). We used three bacterial concentrations whose measured optical densities (ODs) were 0 (control), 0.05 (low concentration), and 0.31 (high concentration), respectively. For the experiments with bacterial colonies, cells were grown on agar plates. A plastic inoculating loop was dipped into bacterial suspensions, and the inoculating loop was immediately streaked gently over a quarter of the agar plate using a back and forth motion. The agar plates were incubated at in 37°C incubator for 12 h to form bacterial colonies.

**Chicken breast tissue preparation**

Fresh chicken breast meat was purchased at a local food market on the date of the experiments. To make samples with different levels of contamination, cultured bacterial suspensions at various concentrations were added to the fresh chicken breast meats which were then incubated in a 37°C incubator for 30 minutes. Before measuring the laser speckle images, the chicken breast meat was cut into 2 mm thick slices and then sandwiched between two slide glasses to prevent movements of the samples and drying of water.

**Results and Discussion**

A schematic of the present method is shown in Figure 1a. When a coherent laser beam impinges, reflected light from tissues without living microorganisms exhibit static speckle patterns. Although reflected light has highly complex intensity patterns due to multiply scattered light paths in tissues, these speckle patterns do not vary over time because of the deterministic nature of multiple light scattering. However, the presence of living microorganisms dynamically perturbs light paths in tissues resulting in varying speckle patterns over time. Thus, by measuring and analyzing the dynamic speckle intensity patterns from meat products, one can detect the presence of living microorganisms. Measuring speckle intensity patterns can be achieved with a simple laser illumination and imaging system shown in Figure 1b. Representative speckle intensity images of tissues with and without microorganisms are shown in Figure 1c. It is clearly seen that the uncontaminated tissues have consistent speckle intensity patterns over time. However, tissues with microorganisms have time-varying speckle patterns. The tissues with microorganisms have highly fluctuating intensity values over time (Figure 1d).

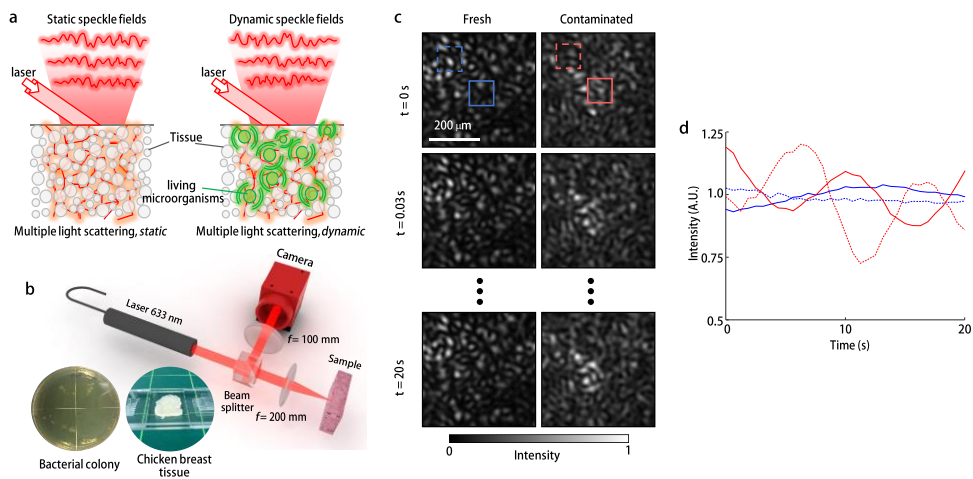

**Figure 1** (a) Schematic showing the detection of microorganisms in meat using time-lapse laser speckle images. In the absence of microorganisms, reflected light from tissue samples have static speckle patterns. However, the presence of microorganisms results in time-varying speckle patterns. (b) Experimental setup. A sample is illuminated with a coherent laser beam. Scattered light reflected from a sample is captured by a CCD camera. (*inset*) Photographs of an agar plate with bacteria colonies and a chicken breast tissue sample. (c) Representative time-lapse laser speckle images. Scale bar is 200 μm. (d) Normalized laser speckle intensities over time obtained in the boxes in (c).

These time-varying speckle signals can be quantitatively addressed with the speckle correlation time[13]. Meat contaminated with living microorganisms have a relatively short correlation time compared to fresh food because of the spontaneous movements of microorganisms. By measuring the correlation time of scattered light from samples, the presence and activity of microorganisms can be quantitatively analyzed.

Previously, the intensity autocorrelation function, $g_2(\tau)$, has been extensively utilized to investigate the movements of scatters in

Brownian motions by relating them to the dynamic speckle patterns, and this technique is also known as diffuse-wave spectroscopy.[22,23]. For the purpose of direct comparison between fresh and contaminated samples, we use the normalized autocorrelation intensity function $C(x,y;\tau)$, which is given by

$$C(x, y; \tau) = \frac{1}{T-\tau} \sum_{t=1}^{T-\tau} \bar{I}(x, y; t)\bar{I}(x, y; t+\tau)\delta t , \qquad (1)$$

where $\bar{I}(x,y;t)$ is the normalized intensity image of the speckle patterns captured at time $t$; $T$ is the total acquisition time; $\delta t$ is the time difference, and $\tau$ indicates the lag time. $(x, y)$ indicates the 2-D spatial coordinates of an image. The normalized autocorrelation intensity function $C(x,y;\tau)$ is exactly identical to the normalization of $g_2(\tau) - 1$.

To demonstrate that the spontaneous movements of food-borne microorganisms in tissue can be detected by measuring and analyzing laser speckle patterns reflected from the tissue, we did experiments with cultured bacterial colonies on agar plates. The speckle intensity patterns from the agar plates in the absence of bacterial colonies and presence of two different bacterial species (*E. coli* and *B. cereus*) are shown in Figure 2a. Due to significantly inhomogeneous RI distributions in the agar plates, the illumination of coherent laser beams results in speckle formations in the reflected light for all cases. The speckle pattern taken after 10 secs for the agar plate only shows almost no change, indicating that the scatters do not change over time. However, the agar plates with bacterial colonies exhibit a significant change in the speckle patterns regardless of the bacterial species (see also Supplementary Movie 1). This result implies that the movements of live bacteria in agar scramble the interference conditions for the speckle patterns. Regarding optical path lengths, the displacement of fractions of the laser wavelength is large enough to change the speckle patterns significantly.

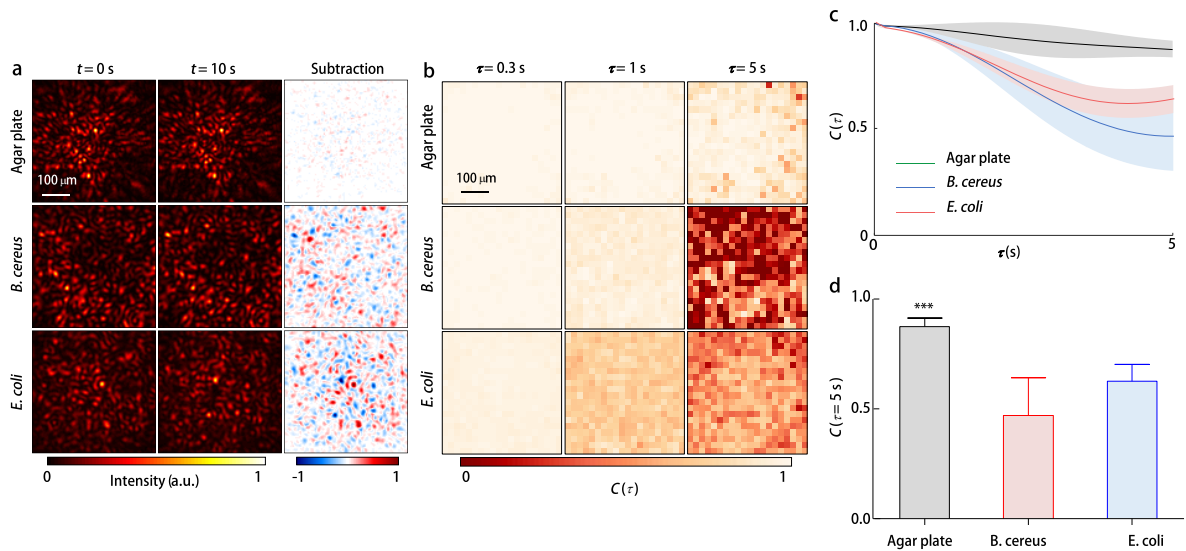

**Figure 2** Laser speckle patterns obtained from agar plates with and without bacterial colonies. (a) Representative laser speckle images obtained at $t = 0$ s and 10 s are shown in the left panel. Subtraction of two laser speckle images at $t = 0$ s and 10 s is displayed on the right panel. Scale bar is 100 μm. (b) Representative autocorrelation function, $C(\tau)$, maps of speckle patterns calculated at $\tau = 0.3$ s, 1 s, and 5 s. (c) Averaged values over the entire area. The shed areas indicate standard deviation ($n = 5$) (d) Quantification of averaged $C(\tau)$ values at $\tau = 5$ s. The symbol *** indicates p-value < 0.001 for comparison among samples using ANOVA followed by posthoc Tukey's method ($n = 5$).

For quantitative and efficient analysis of the time-varying speckle intensity patterns, the intensity autocorrelation maps $C(\tau)$ are shown in Figure. 2b for all three cases. The agar plate without bacterial colonies had almost constant values of the autocorrelation over space and time. However, the agar plates with *E. coli* and *B. cereus* had decreases in the autocorrelation values over time. Interestingly, the amount of decrease in the autocorrelation values was different for *E. coli* and *B. cereus* (Figures 2b–c). This deviation can result from different movements and RI values of the bacterial samples.

To further analyze the autocorrelation maps, averaged autocorrelation maps over the entire area as a function of the time lag are shown in Figure 2c. Over a short time lag ($\tau < 0.5$ s), few changes were observed regardless of the presence of bacteria. However, for a longer time lag ($\tau > 1$ s), the agar plates with bacterial colonies had substantial decreases in the autocorrelation, i.e., temporal dynamics. For the agar plate without bacterial colonies, the correlation value decreased very slowly and monotonically over the time lag, which can be attributed to the mechanical instability of the optical system and water evaporating from the sample . However, the agar plates with bacterial colonies had significant and distinct decreases in the autocorrelation values over the time lag. Autocorrelation coefficients at $\tau = 5$ s obtained from the agar plates without bacteria and with *E. coli* and *B. cereus* were $0.88 \pm 0.04$, $0.47 \pm 0.17$, and $0.63 \pm 0.08$, respectively (Figure 2d). The decreases in the autocorrelation at $\tau = 5$ s are statistically different compared to the control case (*p*-values < 0.001). These results indicate that laser speckle decorrelation is proportional to the bacterial activity. Moreover, there is a similar tendency between the experimental results for the two bacterial strains which suggests that the present method can be used independent of bacterial strains.

To demonstrate the applicability of the present method for investigating the level of food contamination, we performed experiments using chicken breast meat. The representative correlation maps obtained from the meats treated with a PBS solution (OD = 0), low concentration of *B. cereus* (OD = 0.05), and high concentration of *B. cereus* (OD = 0.32) are shown in Figure 3a. The meats contaminated with bacteria had significant decreases in the autocorrelation values over the time lag, whereas the control group (the meat treated with a PBS solution) did not show any major changes. PBS solution was added to the control group to exclude the possibility that the evaporation of water may cause fluctuations in the speckle patterns over time. It is noteworthy that the meat treated with a high concentration of bacteria had more significant changes over the time lag compared with the meat treated with a low concentration of bacteria (Figures 3a-b). Furthermore, the amount of decreases in the autocorrelation values are proportional to the concentration of the treated bacteria. The measured autocorrelation values for the function $C(\tau)$ at $\tau$ = 10 s were 0.94 ± 0.01, 0.65 ± 0.33, and 0.31 ± 0.12 for the PBS solution, low concentration, and high concentration of bacteria, respectively (Figure 3c). The autocorrelation values are statistically different between the groups ($p$-values < 0.001), and the decreases in the autocorrelation are proportional to the concentration of bacteria. These results show that the analysis of laser speckle decorrelation enables the measurement of the level of food contamination caused by microorganisms.

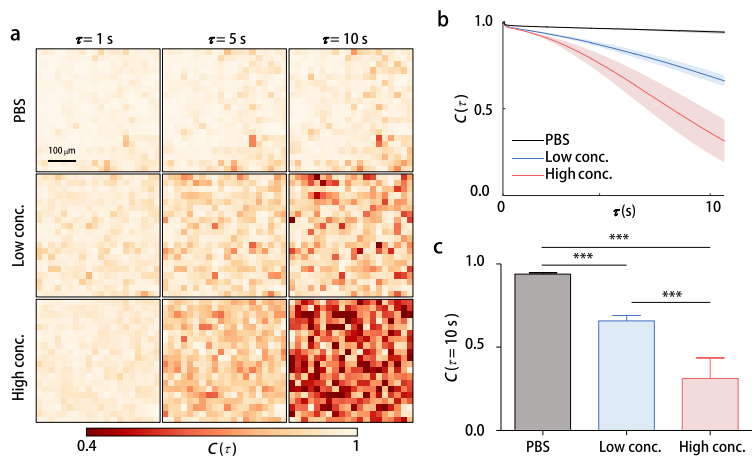

**Figure 3** Assessing bacterial activity in meat. (a) Representative autocorrelation maps in meat treated with various concentrations of bacteria at various time lags. (b) Averaged $C(\tau)$ values over the areas in (a) as a function of the time lag. The shed areas indicate standard deviation ($n$ = 5) (c) Quantification of the autocorrelation values at $\tau$ = 10 s. The symbol *** indicates a p-value < 0.001 for comparison among the groups using ANOVA followed by posthoc Tukey's method. ($n$ = 5)

Next, we further tested the relationship between the bacterial activity in meat and the speckle intensity decorrelation. To suppress bacterial activity, we exposed the meat contaminated with the high concentration of bacteria to ultraviolet (UV) light radiation. UV radiation is widely used for sterilization due to its antibacterial effects. [24,25] The fresh chicken breast meat was dipped into a cultured bacterial suspension (OD = 0.32) and incubated for 30 minutes in a 37°C incubator. Then, the contaminated chicken breast meat was irradiated with a UV lamp (30 W) in a clean bio-hood bench (JSCB-1200SB, JS Research Inc., South Korea).

The effects of the UV irradiation on the autocorrelation function are shown in Figure 4. The autocorrelation maps at $\tau$ = 10 s obtained with the contaminated chicken meats exposed to various UV doses are shown in Figure 4a. It is clearly seen that the contaminated meat unexposed to UV radiation had strong decorrelation in the speckle dynamics, implying vigorous bacterial activity. However, UV radiation for 1 min. resulted in significant increases in the autocorrelation values indicating the suppression of bacterial movements. A longer UV radiation period caused further increases in autocorrelation values, showing a significant suppression of the bacterial activity. The measured autocorrelation coefficients for $\tau$ = 10 s were 0.23 ± 0.10, 0.73 ± 0.08, 0.84 ± 0.03, and 0.91 ± 0.01 for no UV radiation and UV radiation for 1, 5, and 15 min., respectively (Figure. 4c). This result clearly demonstrates the decorrelation in laser speckle dynamics as a result of the bacterial activity. Additionally, it also shows that the present method can be effectively used for quantitative, non-contact, and non-invasive assessment of bacterial activities in meat.

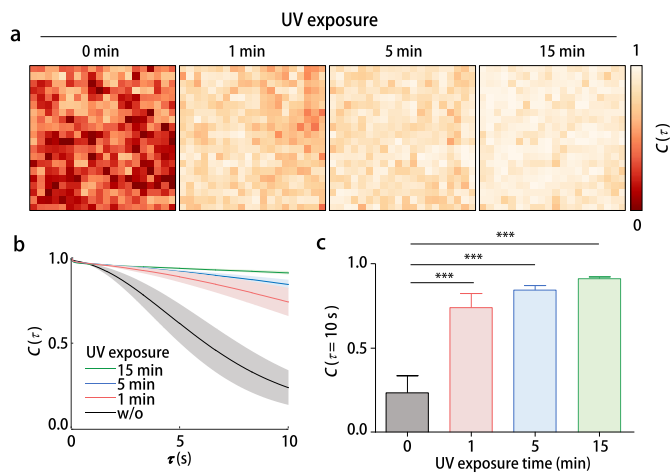

**Figure 4** The relation between bacterial activity in chicken breast meat and laser-induced speckle decorrelation shown by exposing

meat to UV light. (a) Representative autocorrelation maps at $\tau = 10$ s of meat with various UV exposure times. (b) Averaged $C(\tau)$ values over the entire areas in (a) as a function of the time lag. The shed areas indicate standard deviation ($n = 5$) (c) Quantification of averaged autocorrelation values at $\tau = 10$ s. The symbol *** indicates a p-value < 0.001 for comparison among groups using ANOVA followed by posthoc Tukey's method. ($n = 5$)

**Conclusion**

In conclusion, we proposed and experimentally demonstrated a simple but powerful laser technique for assessing bacterial activity in food by analyzing the laser speckle decorrelation. Through various experimental validations, we show that spontaneous bacterial activity causes strong decorrelation in laser speckle dynamics. Additionally, we demonstrate that the present method can also be applicable to various bacterial strains and base media.

The present method has several advantages. First, this method is non-contact and non-invasive. Unlike other conventional chemical or molecule methods which inevitably involve invasive and contact procedures, the present method is based on the analysis of dynamic laser speckles which can be obtained by simply measuring the reflectance of a laser beam from a sample. Meats sealed with transparent plastic wraps can also be examined with the present method because the present method only detects time-varying signals in reflected laser beams and a transparent plastic wrap does not cause time-varying signals. Second, the present method can provide rapid assessment. The activity of live bacteria can be identified within a few seconds. Third, the technique is extremely simple and cost-effective. Without precise optical alignment, using a simple instrument consisting of a coherent laser source and an image sensor can achieve the non-invasive assessment of bacterial activity in food. The instrumentational simplicity garners significant flexibility in its applications. For example, the present method can be integrated as a compact optical module, and it can be implemented in a refrigerator in homes. The present method can also be used in the food industry; a laser system can be implemented in food manufacturing or storage line. The present method also has huge potential for developing countries with a lack of biochemical laboratory facilities. Fourth, the present method can used for various types of bacterial strains and species. Because the principle of this method is based on the alterations in the laser self-interference due to the dynamic movements or activities of live bacteria, no prior information about the bacteria is necessary to detect bacterial activities in food.

Although the present method can be effectively used for the detection of bacterial activity, it is not able to identify different pathogenic bacterial strains, such as *Salmonella*, *Listeria*, *B. cereus*, *E. coli*, and *Campylobacter*.[21] Nonetheless, the present method can be potentially used for various applications where the detection of bacterial activity is more important to avoid food toxicity or to perform prescreening tests. In addition, the present method can also be exploited in research applications where the quantification of bacterial activities or the efficacy of antibacterial drugs is involved. Therefore, the present method can be a versatile tool for assessing bacterial activities in numerous applications and fields due to its simple, non-destructive, and rapid procedures.

**Acknowledgments**

This work was supported by KAIST, and the National Research Foundation of Korea (2015R1A3A2066550, 2014K1A3A1A09063027, 2012-M3C1A1-048860, 2014M3C1A3052537) and Innopolis Foundation (A2015DD126). We thank Dr. Seungwook Ryu for his help with the experiments.